 \documentclass[page-classic]{epl2} %for one column style

\title{Divergent Vortex Mass in a Superconducting Film in Proximity to a Metal}
\author{D. M. Gaitonde}
\institute{                    
   Human Resource Development Division, Bhabha Atomic Research Centre, Mumbai 400085, India
 
}

\abstract{We consider a moving vortex in a two dimensional superconductor located at a distance $d$ from a metallic overlayer. Starting from the microscopic imaginary time action we integrate out the electronic degrees of freedom to obtain a low energy, long wavelength effective action for the vortex. We focus our attention on the vortex kinetic energy and derive a general expression for the vortex mass. We find that in the limit $d\rightarrow \infty$ the Coulomb screening of the density fluctuations, associated with vortex motion, results in a very small vortex mass as has been obtained in earlier studies. In the opposite limit of $d\ll \xi $ where $\xi$ is the coherence length of the superconductor we find that the vortex mass diverges logarithmically with the size of the system as the proximity to the metal makes the screening processes, that usually make the mass small, ineffective.
We comment on the relevance of our results to recent experiments which show a dramatic fall in resistance when a metallic gate is placed near a supeconducting film in a magnetic field at low temperature.}

\begin{document}

\maketitle

%\section{Section title}
The vortex mass is a basic parameter in studies of vortex dynamics. Over the years there have been several estimates~\cite{Suhl,Coffey, Leggett,DMG1,Blatter,Ao,Kopnin} of its size which have often disagreed with each other. In recent times there has been renewed interest in this question because of the possibility of new phenomena involving quantum dynamics of vortices such as quantum flux creep~\cite{qcreep} and quantum melting of the vortex lattice~\cite{qmelt}. While the precise role of the mass remains unclear as the vortex dynamics at low temperature is complicated due to the presence of dissipation as well as the Magnus force
it seems intuitively plausible that the size of the mass is a rough measure of the importance of quantum effects in describing vortex dynamics.

In this letter we consider a two dimensional superconductor in proximity to a metallic overlayer that is seperated from the superconductor by a distance $d$. We consider a moving vortex in the superconductor whose
instantaneous position is ${\bf R_0(\tau)}$ where $\tau$ is the imaginary time variable. We consider a phase-only approach in describing the vortex which is valid at length scales larger than the coherence length of the superconductor $\xi$. We start with the microscopic imaginary time action  and derive the low energy, long wavelength effective action for the system. We had earlier derived the effective action for such a system~\cite{DMG2}. However the focus of that work was the non-singular longitudinal phase fluctuations in the absence of a magnetic field whereas the subject of the present work is a transverse vortex configuration of the phase of the superconducting order
parameter. Having obtained the effective action in terms of the vortex co-ordinate and the vortex velocity we proceed to solve the equations of motion for the three dimensional Coulomb potential $ A_0({\bf r},z,\tau)$. Substituting the solutions thus obtained for the Coulomb potential in the action we obtain the vortex kinetic energy.

Examining the co-efficient of the vortex  kinetic energy we are able to write an expression for the vortex mass. We explicitly evaluate the mass in two different limits. In the first case we consider the limit $d\rightarrow \infty$ which corresponds to the situation where the metal is absent.
In this case we find the mass from the far region to be negligibly small because of efficient screening in agreement with earlier studies~\cite{Suhl,Leggett,DMG1}. Of course in this case the true mass is somewhat larger and to evaluate it one has to  consider the contribution coming from the core of the vortex~\cite{DMG1,Blatter,Ao,Kopnin}. While the precise value of this contribution remains controversial, the important point from the point of view of this work is that it is finite. We then consider the value of the vortex mass in the limit  $d\ll \xi $ where $\xi$ is the superconducting coherence length. We find that the presence of the metal destroys the screening observed in the first case and the mass is divergent. In order to get a finite answer, we introduce a long distance cutoff $R_c$ which corresponds to the system size. We then find that the mass scales as $\ln(R_c/\xi)$. 

The big change in the vortex mass with and without the metal being present has direct experimental significance. Mason and Kapitulnik~\cite{Kapitulnik} have carried out measurements of the electrical resistance in amorphous superconducting films in the presence of a magnetic field. They carried out their measurements on two types of samples: a) with a conducting ground plane at a distance d from the sample
b)without the conducting ground plane. They found that at low temperature ($T\rightarrow 0$) there is a levelling off of the resistance to a finite value indicative of a metallic phase in samples without the conducting ground plane. The introduction of the ground plane inhibits the resistance levelling and instead causes a sizable decrease in the value of the resistance. These experiments were recently interpreted by Michaeli and Finkel'stein~\cite{Fink1,Fink2} as evidence for presence and absence of vortex tunneling in the two cases. They argue that the magnetic coupling between the vortices in the superconducting film and the electrons in the conducting ground plane inhibits the tunneling of vortices. In addition to the effects considered by them another important factor contributing to the suppression of vortex tunneling in the samples with a conducting plane placed near the superconductor is the dramatic increase in the vortex mass in this case. It is to be noted that the studies of Mason and Kapitulnik~\cite{Kapitulnik} were carried out for $d\approx 160 \AA$ with $\xi \approx 250 \AA$ which is reasonably described by our calculation. 

We now turn to the details of our calculation. The dynamics of the coupled electronic subsystems is described by the action
\begin{equation}
S=\int^{\beta}_{0}d\tau\int d^{2}r\int dz[\mathcal{L}_{sc}+\mathcal{L}_{eg}+\mathcal{L}_{em}+\mathcal{L}_{ion}]
\end{equation} 
where
\begin{equation}
\mathcal{L}_{sc}=\delta (z)[\sum_{\sigma} \overline{\psi}_{\sigma}({\bf r},\tau)({\partial\over \partial\tau} +h_{sc}){\psi}_{\sigma}({\bf r},\tau)
+{\mid\Delta({\bf r},\tau)\mid^{2}\over g}+(\Delta({\bf r},\tau)\overline{\psi}_{\uparrow}\overline{\psi}_{\downarrow}+h.c.)]
\end{equation}
\begin{equation}
\mathcal{L}_{eg}=\delta (z-d)[\sum_{\sigma} \overline{\chi}_{\sigma}({\bf r},\tau)({\partial\over \partial\tau} +h_{eg}){\chi}_{\sigma}({\bf r},\tau)]
\end{equation}
\begin{equation}
\mathcal{L}_{em}={[\nabla A_0({\bf r},z,\tau)]^2\over 8\pi} +{[\nabla \times {\bf A}({\bf r},z,\tau)]^2\over 8\pi}
\end{equation}
and 
\begin{equation}
\mathcal{L}_{ion}= ieA_0({\bf r},0,\tau)\overline\rho_{\psi}\delta(z)+ieA_0({\bf r},d,\tau)\overline\rho_{\chi}\delta(z-d)
\end{equation}
The electrons at $({\bf r},\tau)$ with spin $\sigma$ are represented by the Grassman field variables $\overline{\psi}_{\sigma}({\bf r},\tau)$, ${\psi}_{\sigma}({\bf r},\tau)$ and $\overline{\chi}_{\sigma}({\bf r},\tau)$, ${\chi}_{\sigma}({\bf r},\tau)$ in the superconducting layer (at $z=0$) and the electron gas (at $z=d$) respectively. Here 
\begin{equation}
h_{sc}={(-i\hbar\nabla_{\mid\mid}-e/c{\bf A}({\bf r},0,\tau))^2\over 2m_{sc}} -ieA_0({\bf r},0,\tau)+V_{sc}({\bf r}) -\epsilon_F^{sc}
\end{equation}
and 
\begin{equation}
h_{eg}={(-i\hbar\nabla_{\mid\mid}-e/c{\bf A}({\bf r},d,\tau))^2\over 2m_{eg}} -ieA_0({\bf r},d,\tau)+V_{eg}({\bf r}) -\epsilon_F^{eg}
\end{equation}
Thus, $\mathcal{L}_{sc}$ includes the electronic kinetic energy and the coupling of the superconducting electrons at $z=0$ to the electromagnetic potentials as well as to a random potential. The field $\Delta$ is the auxilliary Hubbard-Stratonovich field obtained from the BCS contact interaction and g is the strength of the attractive interaction. $\mathcal{L}_{eg}$ describes the two dimensional electron gas at $z=d$ together with its coupling to a random potential $V_{eg}$ and the electromagnetic potentials. $\mathcal{L}_{em}$ gives the electric and magnetic field energies of the system. $\mathcal{L}_{ion}$ describes the interaction of the Coulomb potential with neutralizing positively charged ionic backgrounds. 

We consider an order parameter configuration that corresponds to a uniformly
moving vortex whose instantaneous position is ${\bf R}_0(\tau)$. Restricting our attention to the ''far region'' outside the vortex core
we ignore the spatial dependence of the amplitude of the order parameter and make the replacement $\Delta({\bf r},\tau)=\Delta_0\exp [i\phi({\bf r}-{\bf R}_0(\tau))]$ where $\nabla_{\mid\mid}\phi=\widehat{z}\times {{\bf r}-{\bf R}_0(\tau)\over \mid {\bf r}-{\bf R}_0(\tau)\mid^2}$. Then on going to a gauge in which the order parameter is real~\cite{Ambegaokar,TVR} (i.e. making the transformation $\psi\rightarrow \exp[{i\phi/2}]\psi$)and then integrating out the electrons, both in the superconducting layer and the metallic layer,
we obtain at low energies and long wavelengths the effective action for the system to be given by
\begin{equation}
S_{eff}=\int^{\beta}_{0} d\tau\int d^{2}r\int dz[\mathcal{L}_{M}+\mathcal{L}_{K}+\mathcal{L}_{S}]
\end{equation}
where
\begin{equation}
 \mathcal{L}_{M}=-{i\overline\rho_{\psi}\over 2}\delta(z) \frac{\partial {\bf R}_0}{\partial\tau}\cdot \nabla_{\mid\mid}\phi({\bf r}-{\bf R}_0(\tau))
 \end{equation}
 \begin{equation}
 \mathcal{L}_{K}=[ \delta(z){P_{sc}\over 8}(\frac{\partial {\bf R}_0}{\partial\tau}\cdot \nabla_{\mid\mid}\phi +2eA_0({\bf r},0,\tau))^2+\delta(z-d){P_{eg}e^2\over 2}A_0^2({\bf r},d,\tau)+
 {\nabla A_0({\bf r},z,\tau)^2\over 8\pi}]
 \end{equation}
 and
 \begin{equation}
\mathcal{L}_{S}=[\delta(z){D_{sc}\over 2m_{sc}}({\hbar\nabla_{\mid\mid}\phi\over 2}-e/c{\bf A}({\bf r},0,\tau))^2+ {[\nabla \times {\bf A}({\bf r},z,\tau)]^2\over 8\pi}]
\end{equation}

$\mathcal{L}_{M}$ is the term that leads to the Magnus force on moving vortices and its co-efficient is known~\cite{DMG3,Larkin} to be proportional to the density of electrons in the superconducting layer. We will not discuss this term any further and merely list it for completeness. 
$\mathcal{L}_{S}$ is the standard term for the static energy of the vortex with $D_{sc}$ being the superfluid density of the electrons in the superconducting layer and we will not consider it any further. However, it is worth pointing out that the presence of the metal doesn't lead to any change in the static energy of the vortex at this level of approximation because the superfluid density of the normal metal vanishes in the low energy, long wavelength limit and thus there is no contribution of a term quadratic in the vector potential seen by the electron gas at $z=d$.

We now turn our attention to $\mathcal{L}_{K}$ which corresponds to the vortex kinetic energy. The co-efficients $P_{sc}$ and $P_{eg}$ correspond 
to the ${\bf q}=0$, $\nu_m=0$ limit of the electronic density-density correlation function calculated in the presence of a uniform superconducting gap in the presence of a random potential for the former case and for a two dimensional fermion gas in the presence of a random potential for the latter case. While it is possible to microscopically calculate these co-efficients we will make no attempt to do so but instead
re-express these co-efficients~\cite{Leggett} in terms of the Thomas-Fermi screening lengths in the superconducting and normal layers respectively.

To proceed further, we solve the equations of motion obtained by varying
$S_{eff}$ with respect to $A_0({\bf r},z)$. Varying $S_{eff}$ with respect to $A_0$ we find its equation of motion to be given by
\begin{equation}
\label{eq.12}
\frac{\nabla^2A_0({\bf r},z,\tau)}{4\pi}=\frac{eP_{sc}\delta(z)}{2}[\frac{\partial {\bf R}_0}{\partial\tau}\cdot \nabla_{\mid\mid}\phi({\bf r}-{\bf R}_0) +2eA_0({\bf r},0,\tau)]+P_{eg}e^2\delta(z-d)A_0({\bf r},d,\tau)
\end{equation}
We solve eq. (12) by taking Fourier transforms. We merely quote the final results. 

Defining $A_0({\bf q},z=0,\tau)$ to be the two dimensional Fourier
transform of $A_0({\bf r},z=0,\tau)$ we find its value to be given
by 
\begin{equation}
A_0({\bf q},z=0,\tau)=\frac{-2\pi^2eP_{sc}}{iq^3}\frac{F_1(q)}{F_2(q)}
\exp{(-i{\bf q}\cdot {\bf R}_0)}\frac{\partial {\bf R}_0}{\partial\tau}\cdot \widehat{z}\times {\bf q} 
\end{equation}
 where
 \begin{equation}
 F_1(q)=1+\frac{2\pi e^2 P_{eg}}{q}(1-\exp{(-2qd)})
 \end{equation}
 and
 \begin{equation}
 F_2(q)=1+\frac{2\pi e^2 P_{eg}}{q}+\frac{2\pi e^2 P_{sc}}{q}+\frac{4\pi^2 e^4 P_{eg}P_{sc}}{q^2}(1-\exp{(-2qd)})
 \end{equation}
 The form of $F_1(q)$ and $F_2(q)$ make it apparent that the co-efficientts $P_{sc}$ and $P_{eg}$ are related to the Thomas-Fermi screening lengths in the superconductor ($\lambda_{TF}^{sc}$) and the metal ($\lambda_{TF}^{eg}$) respectively by the relations
 $2\pi e^2P_{sc}=1/{\lambda_{TF}^{sc}}$ and $2\pi e^2P_{eg}=1/{\lambda_{TF}^{eg}}$.
 
 We are now ready to derive an expression for the vortex mass. The vortex kinetic energy can be simplified by substituting the equation of motion (eq. (12)) in  $\mathcal{L}_{K}$ (eq.(10)). We then obtain
 \begin{equation}
 \mathcal{L}_{K}= \delta(z){P_{sc}\over 8}(\frac{\partial {\bf R}_0}{\partial\tau}\cdot \nabla_{\mid\mid}\phi +2eA_0({\bf r},0,\tau))
 \frac{\partial {\bf R}_0}{\partial\tau}\cdot \nabla_{\mid\mid}\phi
 \end{equation}
 Going over to Fourier space with respect to the two dimensional co-ordinate ${\bf r}$ and substituting the result for $A_0({\bf q},z=0,\tau)$ in eq. (16) we can find the vortex mass $m_{vort}$ to be given by
 \begin{equation}
 m_{vort}= \frac{m_{el}}{8}\frac{a_0}{\lambda_{TF}^{sc}}\int_{R_c^{-1}}^{\xi^{-1}}
 \frac{dq}{q}\frac{1+1/(q\lambda_{TF}^{eg})}{F_2(q)}
 \end{equation}
 
 Here $m_{el}$ is the electron mass and $a_0=\frac{\hbar^2}{m_{el}e^2}$ is the first Bohr radius. We have cut off the momentum integration at low momenta at a scale which is the inverse of the size of the two dimensional 
 superconducting film and at high momenta at the scale which corresponds to the inverse coherence length. The result contained in eq. (17)constitutes the main result of this paper. While an analytic evaluation of the integral in eq. (17) is not possible for arbitrary d, useful progress can be made by evaluating it exactly in the limits $d\gg R_c$ and $d\ll \xi$. 
 
 We first condider the case $d\gg R_c$ which corresponds to the situation of a superconducting film without any metallic overlayer. In this case we
 can explicitly evaluate the expression in eq. (17) and find that it is
 given by
 \begin{equation}
 m_{vort}\approx \frac{m_{el}}{8}\frac{a_0}{\xi}
 \end{equation}
 This is a well-known result~\cite{Suhl,Leggett} that the contribution to the mass from the far region is negligibly small. We wish to emphasize that
 the true mass in this case arises from transitions induced in the bound states in the vortex core by the vortex motion and is somewhat larger. While its precise value is controversial~\cite{DMG1,Blatter,Kopnin} the important point from the point of view of this letter is that it is finite
 and relatively small.
 
 We now turn our attention to the other limit $d\ll\xi$. In this case we proceed further by making the approximation $(1-\exp{(-2qd)})\approx 2qd$.
 On making this substitution we can evaluate the integral in eq. (17) and find
 \begin{equation}
 m_{vort}\approx \frac{m_{el}}{8}\frac{a_0}{\lambda_{TF}^{eg}+\lambda_{TF}^{sc}+2d}\ln{(R_c/\xi)}
 \end{equation}
 Thus we find that when the metallic layer is brought close to the superconducting layer, screening which had made the vortex mass small in the absence of the metallic layer is now rendered ineffective. 
 
 These results indicate that as the metallic layer is brought in from large
 distances the vortex mass will continously increase till it becomes
 divergent for $d\ll \xi$. As the vortex mass is a basic parameter that
 has an important bearing on the importance of quantum effects in vortex dynamics, our result suggests that the strength of quantum effects can be manipulated by bringing a metallic layer close to the superconducting film.
 In particular one can conceive of a metal-insulator transition of vortices
 being driven by the absence or presence of a metallic overlayer. As was stated earlier the results of Mason and Kapitulnik~\cite{Kapitulnik} have been previously interpreted~\cite{Fink1,Fink2} as being a realization of such a vortex metal-insulator transition. Our results provide an alternative, Coulomb screening driven, mechanism of this transition. However a quantitative interpretation of these experiments is complicated by the fact that in addition to the vortex kinetic energy there are also the Magnus force and the viscous drag on the motion of the vortex arising from the core~\cite{Blatter,Volovik}. As the size of these effects is uncertain it is difficult to make quantitative estimates of the vortex tunneling rates based on our results for the vortex mass.
 
 Finally let us recapitulate the main points of this letter. We have considered a superconducting film in proximity to a metallic overlayer. Starting from the electronic action for the system we derive the effective action describing the dynamics of a moving vortex in the superconducting layer. We find that the presence of the metallic layer makes the vortex mass
 divergently large. This result is in agreement with electrical resistivity measurements in a superconducting film in proximity to a conducting ground plane.

\acknowledgments
I wish to thank Dr. R. R. Puri for his encouragement in completing this work.

\end{document}